# A new organizational metrics: A quantum approach

W.F. Lawless
Paine College
1235 15th Street
Augusta, GA 30901-3182
706-821-8284 office
706-664-8148 cell
lawlessw@mail.paine.edu
homepage.mac.com/lawlessw

Joseph Wood, LTC
US Army
Ft. Gordon, GA 30905
706-787-2875
Joseph.C.Wood@usarmy.mil

Lily Tung
Paine College
1235 15th Street
Augusta, GA 30901-3182
706-396-7594 office
tungh@mail.paine.edu

**Abstract:**
A future goal of robot teams and agent-based models (ABMs) is to field organizations and systems based on first principles derived from human counterparts. Forestalling that opportunity, the failure of traditional organizational theory has at the same time opened the way to innovative theories of organizations and change. Inspired by Bohr and Heisenberg about the application of interdependent uncertainty in the interaction between action and observation, making organizations bistable, we have begun to construct a theory of organizations based on the uncertainty of energy level (resources) and belief/action consensus, leading to preliminary metrics of organizational performance that we have discovered in field studies. Our goal in this project is to address the problem posed by organizations with: the development of new theory; field tests of new metrics for organizations; and the development of quantum ABMs set within a social circuit as a building block for an organization. Should we be successful, our research would represent a fundamental departure from traditional observational methods of social science by forming the basis of a predictive science of organizations. We expect that replacing the traditional method of observation with a predictive science must account for when cognitive observations work and when they do not (illusions).

**Introduction:**
Conant and Ashby (1970) proposed that "Every good regulator of a system must be a model of that system". Yet the "discontent with existing theoretical perspectives" (Schoonhover et al., 2005, p. 453) indicates that the evidence for traditional models of organizations fails to meet that criterion: The correspondence presumed between reality and observations carefully aggregated from individual members has been unable to reconstruct an organization's actual status (Levine & Moreland, 1998). The result clearly indicates that the traditional model cannot be used to predict or control organizational outcomes (Weick & Quinn, 1999). Indeed, over the years, theory for organizations has become more rather than less fragmentary (Miner, 1982; 2002). This failure has led Pfeffer & Fong (2005) to propose that belief illusions are a critical missing ingredient. We agree, and have constructed a bistable theory of reality interdependent between physical reality and observations. Observations can be modeled with complex functions where only the real part corresponds with reality while the imaginary part does not (i.e., "illusions"). If the imaginary part is dependent on social-psychological influences (e.g., culture, roles) spread across a field of observations underpinning multiple perspectives embedded in physical space (Lewin, 1951), two incommensurable stories of the same social reality always exist (Wendt, 2005).

The immediate implication of accepting the existence of illusions and multiple perspectives is that the traditional reliance on self-reported or observational information alone is insufficient to model or control organizations. Yet, by proposing this research project, we endorse the belief of Answorth and Carley (2007, p. 102) that computational organizational science has an opportunity to contribute to the discovery and validation of theory. Further, a theory of organizations with dual natures, such as Scott's (2004) duality of production and social systems, is not novel. What makes our approach novel is to construct a computational model that does not rely on the simple convergence processes inherent in traditional social science (Campbell, 1996).

From field studies of citizen organizations advising the Department of Energy on its nuclear waste cleanup, we have found that consensus compared to majority rules tend to promote risk perceptions ("illusions"; Lawless & Whitton, 2007). Given that traditional consensus rules are designed to reduce the conflict associated with majority rules (Bradbury et al., 2003), we found instead that consensus rules dampened the underlying motivation to marginalize opposing views as an organization attempts to create a culture around a single story (Atran et al., 2006; viz. reducing entropy). Conflict alone increases information (Myer et al., 2006) while managed conflict improves learning and solution outcomes (Dietz et al., 2003). In the tradeoff between adaptability and innovation, firms that manage the conflict from paradox can simultaneously optimize existing product lines and innovate (Smith & Tushman, 2005). Conflict is driven by opposing views but is managed by having sufficient neutrals on hand to decide an issue (Kirk, 2003); e.g., in 2006, the loss of neutrals presaged the split in the Episcopal Church and the separate conflicts in Palestine and Beirut, but also the loss of market share in consumers deciding whether to buy web advertising from Microsoft or cars from Ford. We have hypothesized that opposing drivers of a decision are able to entangle neutrals into deciding an issue (Lawless et al., 2006b). But more importantly, the entangled system of managing conflict dampens the reliance on illusion.

Organisms live under uncertainty partly dispelled by social interaction (Carley, 2002). To survive, they form organizations as centers of cooperation (Ambrose, 2001) that marginalize opposing beliefs among members in exchange for a share of resources, but in a tradeoff between the loss of information from consensus-seeking and the gain of information from conflict associated with innovation and change. For organizations constituted of bistable agents existing in a consensus field, the loss means that control information must be generated from perturbations (Lawless & Grayson, 2004). Conant and Ashby (1970) have further posited that good regulation occurs when the available control variance is greater that the perturbation variance. In partial support, a recent study at the MIT Media Lab found that professors in the lab produced less entropy than incoming students (Eagle, 2005).

We next make assumptions to "reduce the complexity of the real world" (Budzinski & Christiansen, 2006, p. 29). Assuming that all organizations eventually evolve to an end state of maximum entropy, control occurs in a competition between evolution and importing entropy ("negentropy") that temporarily holds random evolution at bay but also permits an organization, system of organizations or network to become more complex to solve difficult problems (Reyck & Herroelen, 1996; Yu & Efstanthiou, 2002). From May (1973/2001), we assume that during environmentally unstable periods such as 9/11 in the US, evolution is dampened by organizational dynamics, the opposite occurring during environmentally stable times.[1] If true, then in addition to advantages gained from increasing scale by growing or merging, during stable periods, organizations are motivated to gain size to better protect against downturns, but they are precluded from doing so unless sufficient resources are available, including by being more efficient (Andrade et al., 2001).

We have applied our model to construct a web-based metric for Marine Corps weather forecasters (Lawless et al., 2006a); to reorganize a Management Information System at a University in the European Union (EU-MIS; Lawless et al., 2006b); to measure the performance of a military medical department of clinical research (MDRC; Lawless et al., 2006c); and, in an ongoing application, to measure the performance of a central business university's graduate school (CBU; Lawless et al., 2007). This versatile metric derives from applying the quantum model of interdependence to the social interaction (Wendt, 2005), the topic for a symposium in 2007 (www.aaai.org/Symposia/Spring/sss07symposia). The quantum aspect indicates that information entangled among social objects once measured collapses into one of two observables, necessarily losing interdependent information on the non-observed variable.

---

[1] Limited support derives from annual data for August 01 to February 07, producing a correlation between the annual Russell 2000 and VIX indices of -.87, and a t-test of -2.9 significant at $p = .02$, versus the Dow Jones Industrial Average of -.78 and a t-test of -2.49, not significant at $p = .07$; contrasting VIX volatility versus the Dow Jones Industrial and Russell 2000 indices for 12 months post-9/11 gives respective correlations for 2001-02 of -.94 and -.92 with t-tests of -8.54 and -7.42, both significant at $p$ of .000, versus for 2005-06 of -.08 and -.29 with t-tests of -.24 and -.96, both non-significant at $p$ of .81 and .36.

The loss of information opens a new area of study as indicated by the tradeoffs in two very different studies. First, in a meta-analysis of over 30 years of research, Baumeister and his colleagues (2005) found that the self-esteem of individuals was strongly consistent with their worldviews but not with their academic or work performances, compromising the value of self-reports. Then in field studies of the Department of Energy's Citizen Advisory Board (CAB) recommendations on cleaning up nuclear wastes at DOE sites, we have found that decisions by consensus ruled CAB's were rationally consistent but not practical for their DOE sponsor, while decisions by majority ruled CAB's were rationally inconsistent[2] but practical (Lawless & Whitton, 2007). The former result indicates that the problem with consensus is not from arriving at one, but seeking it (Levine & Moreland, 2004). Seeking consensus not only reduces information but it gives inordinate power to select individuals in a group, consequently permitting a group to be more easily controlled by subterfuge,[3] making it a desirable means of governing citizens for autocrats (Kruglanski et al., 2006), but less desirable for organizations (e.g., Unilever has restructured away from dual CEO's, and Shell from dual power centers); in contrast, we have found that majority rules not only generate information but produce as many consensus[4] decisions more quickly and with greater practical value (Lawless et al., 2006b). But while the latter result suggests that organizational outcomes can be controlled, its inconsistency violates the traditional definition of "rational" as normatively consistent (Shafir & LeBoeuf, 2002).

If a fragmented market produces more information, a consolidated market reduces the information available. But in the transition between these extremes, caused for example by a hostile takeover in a fragmented market, controlled perturbations invoke a succession of rapid, complex tradeoffs during the acquisition process that we have termed the measurement problem (Figure 1; from Lawless et al., 2005). Of course, "jolts" that produce discontinuous change arise naturally (Meyer, 1982). In contrast, we have argued that intuitively induced perturbations are used by managers as a tool to test opponents and potential targets in the contest for scale. But in our goal to replace intuition with rational control, it becomes necessary to be able to predict the outcome of these perturbations, invoking the measurement problem.

In Figure 1, uncertainty in the social interaction is represented by an interdependence between strategy, plans, or knowledge uncertainty, $\Delta K$ (where $K$ is a function of the social location where it was learned; from Latané, 1981) and uncertainty in the rate of change in knowledge or its execution as $\Delta v = \Delta(\Delta K/\Delta t)$; this agrees with Levine and Moreland (2004) that as consensus for a concrete plan increases ($\Delta K$ reduces), the ability to execute the plan increases ($\Delta v$ increases). By extension, interdependence also exists in the uncertainty in the resources[5] expended to gain knowledge, $\Delta R$, and by uncertainty in the time it takes

---

[2] Inconsistency derives from summing self-reports over the opposing viewpoints of those who drive a decision as well as those of the neutrals who hold a spectrum of viewpoints; e.g., in legal decision-making about mergers, competing experts with competing models serve to muddy the water (Budzinski & Christiansen, 2006, p. 21).

[3] For example, "The requirement for consensus in the European Council often holds policy-making hostage to national interests in areas which Council should decide by a qualified majority." (WP, 2001, p. 29).

[4] It could be argued that the consensus from majority rules is qualitatively different because it leads to a consensus of action as opposed to the traditional view of consensus in forming a single worldview; we have found that consensus seeking resorts to the least common denominator of a single worldview that is less likely to produce a plan of action (Lawless et al., 2005; Lawless & Whitton, 2007).

[5] Although Bohr (1955) and Heisenberg (1958) believed that uncertainty in atomic interactions could model uncertainty in social interactions (Equations 1 and 2 in Fig. 1), their direct application to the social interaction involves orders of magnitude separating atomic and social levels. But, briefly, at the level of neurons, the eye acts as a quantum information processor that converts photons hitting the retina into usable information (French & Taylor, 1978). Luce (1997) concluded that the Bèkèsy-Stevens quanta model, based on detecting discrete stimulus differences from background along one physical dimension such as $E$ or frequency, $\omega$, to produce a linear relationship between threshold and saturation, works as well as ROC curves in signal detection theory. And from Penrose (Hagan et al., 2002), if the brain can be modeled as a single unit, and if $c$ is Planck's constant, $h$, then converting $\Delta R \Delta t$ into $\Delta E \Delta t = \Delta(\hbar \omega/2\pi)\Delta t \geq h/2\pi$ reduces to $\Delta \omega \Delta t \geq 1$. Choosing $\omega$ as gamma waves in the brain associated with object awareness at about 40 Hz

to enact knowledge, $\Delta t$ (for proofs, see Lawless et al., 2006b). That these two sets of bi-stable factors are interdependent means that a simultaneous exact knowledge of the two factors in either set is precluded.

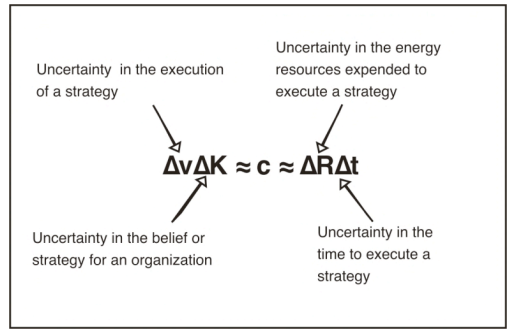

**Figure 1.** Equations 1 and 2: The measurement problem from the perspective of a merger target (bi-sided uncertainty relations exist for the acquiring organization). For example, **Strategy**: after AT&T Wireless put itself on the auction block in 2004 and Cingular made the first offer, AT&T Wireless did not know whether bids would be received from other players such as Vodaphone, or how much more would be offered; **Execution**: Cingular expected that AT&T Wireless would execute its strategy by choosing the best bid by the deadline it had set, an expectation that turned out to be incorrect; **Resources**: AT&T Wireless did not know whether Cingular or Vodaphone would increase their bids to an amount it considered sufficient; **Time**: while the bidders believed incorrectly that the deadline was firmly established, AT&T Wireless was uncertain of the time when the bids would be offered. Finally, although power goes to the winner, it was not easy to determine who won and who lost in this auction. AT&T Wireless was unable to enact phone number portability and became the prey, but its CEO extracted a superior premium for his company and stockholders; while the merger on paper made Cingular the number one wireless company in the U.S., it may have overpaid for the merger; and during the uncertainty of regulatory review (both the length of the regulatory review period and the regulatory decision), with AT&T Wireless losing customers as cable and other competitors exploited the regulatory uncertainty, it was unknown how costly the eventual merger would be based on the assets remaining once the merger had been consummated.

We have also considered the loss of information in business mergers (Lawless & Grayson, 2004). When a market is highly fragmented, like the current U.S. airline industry, it is unable to act cohesively, characterized on average by a loss of profit. In late 2006, US Airways made a hostile offer for Delta Airlines that could have further consolidate the U.S. airline industry, had it been successful. In the first tradeoff, as the average size in market share increases, a more focused business model implies an increase in the ability to execute; a second, related tradeoff occurs that increases the market's capacity on average to put more resources into executing plans more quickly. A more focused market twice the size of a fragmented market should execute in one-half the time (where a focused business model reflects a reduction of organizational duplication, personnel or overhead expenses; or correspondingly, an increase in operational readiness could occur with the wider deployment of new technology).

As an example of applying Eqn. (1) and (2) in Figure 1 to mergers, given that $\Delta R \Delta t = c,$ in a situation where organization-1 is interested in the effects of a merger, all else being equal, then
$$\Delta_1 R \Delta_1 t = c = \Delta_2 R \Delta_2 t. \qquad (3)$$
A merger that leads to an organization with twice the resources it had before becomes:
$$\Delta_1 R = 2\, \Delta_2 R. \qquad (4)$$
Then,
$$\Delta_1 t = c / \Delta_1 R = c / (2\, \Delta_2 R) = c / (2\, c / \Delta_2 t) = \Delta_2 t / 2, \qquad (5)$$

---

gives $\Delta t$ of 25 ms, a reasonable minimum in that awareness occurs between 200 to 500 ms, even though responses to stimuli occur in a much shorter interval but outside of mindful awareness (for supporting brain wave data, see Hagoort, 2004).

meaning, as predicted, that the merged firm can enact its plans twice as fast. Size alone does not mean that the speed of enacting plans will increase unless consensus is also achieved. In the situation where organization-1 has twice the action consensus for its business plan, then,

$$\Delta_1 K = \tfrac{1}{2} \Delta_2 K. \quad (6)$$

Given that

$$\Delta_1 K \Delta_1 v = c = \Delta_2 K \Delta_2 v. \quad (7)$$

Then,

$$\Delta_1 v = c / \Delta_1 K = c / \Delta_2 K / 2 = 2 c / \Delta_2 K = 2 c / c / \Delta_2 v = 2 \Delta_2 v. \quad (8)$$

This means that if organization-1 has ½ the variation in its consensus for action as organization-2, then organization-1 is able to execute twice as fast.

We next apply a rate equation, $\Gamma$, to simplify but further explore the potential to merge (Lawless et al., 2006b):

$$\Gamma = N_1 N_2 v_{1-2} \sigma_{1-2} \exp(-\Delta R_r / \langle R_a T \rangle) \quad (9)$$

where $N_1$ and $N_2$ reflect the internal consensus by the numbers in each group who support the merger proposed by organization-1 to acquire target organization-2; $v_{1-2}$ reflects the information flow from interactions over the merger; $\sigma_{1-2}$ represents resonance between the organizations from a mutual consensus of the merger plan; and the Arrenhius (exponential) term reflects the minimum resources required to overcome barriers to explore possible merger configurations, divided by the resources available to address the barriers. Below $\Delta R_r$ (where $-\Delta R_r = \Delta E - T\Delta S$), predator-prey organizations become non-ergodic; i.e., they are unable to explore the full range of possible merger configurations within an acceptable time horizon accessible to control inputs. Finally, while normally not a factor, in the denominator, $T$ reflects unusual excitation of the participants (e.g., the "frenzy" associated with the recent Equity Office Products merger).

The Arrenhius term in Eqn. (9) estimates the probability of the merger going forward. It indicates whether sufficient resources exist to bring about consensus between the proponents driving a merger and the neutrals engaged by the process (e.g., ceteris paribus, the target is more or less neutral to a contest between potential buyers when both have sufficient funds). Holding the available resources constant (see Table 1 and Fig. 2), the probability of the merger improves with a reduction in the financial barriers to the merger. Conversely, holding these costs to merge constant, the probability of a merger increases with the greater availability of resources that can be directed at the merger.

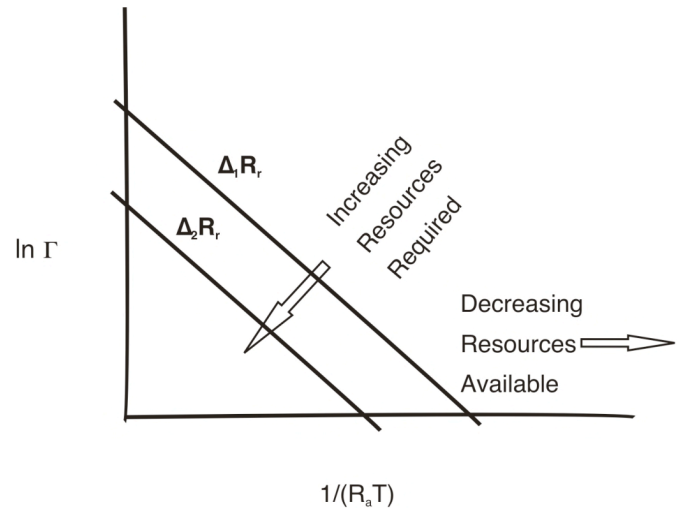

Figure 2. Power law distributions derived from a plot of $\Gamma$ using sample data in Table 1 below.

| $\Delta R_r$ | $\langle R_a \rangle$ | $\Delta R_r$ | $\langle R_a \rangle$ | x | exp (-x) |
|---|---|---|---|---|---|
| 2 | 1 | 2 | 1 | 2 | 0.14 |
| 2 | 2 | 1 | 1 | 1 | 0.37 |
| 2 | 200 | 0 | 1 | 0 | 1.00 |

**Example: The merger of EOP with Blackstone.**

From mid-2006 to its consummation in early 2007, the M&A bidding war between Vornado Reality Trust and Blackstone Group for Equity Office Properties, one of the largest acquisitions ever proposed in the U.S., illustrates Eqn. (9). EOP owned 543 buildings totaling 103.1 million ft$^2$, but, with its stock price at about $45/share, it was considered to be poorly operated, increasing its potential value to the acquirers. And despite VR's larger bid of $41 billion to BG's $38 billion, VR capitulated at the end, suggesting that perception may have been a factor.

Vornado Realty, a well-run firm, bid first for EOP in July, 2006, but was rebuffed in part because of the lengthy time it proposed to close the deal, increasing the uncertainty of its offer. Later, the Blackstone private equity group topped VR's bid in November by 11% with a $36.1 billion all-cash bid that was readily accepted by EOP's Board of Directors in part because BG planned to close the deal quickly. In accepting its bid, BG also had EOP invoke a breakup fee of $200 million. VR countered BG's bid in January 2007 with $52/share followed by BG at $54/share and VR again at $56/share, the latter totaling $41 billion. As the bidding increased, to thwart VR, BG had EOP raise the breakup fee to $500 million and then to $720 million (see also www.equityoffice.com; www.reuters.com ).

In Eqn. (9), $N_1$ and $N_2$ represent internal consensuses. While no apparent issues arose between BG and its partners, there were news reports that VR had internal divisions (Wall Street Journal, 2/2/07, p. C3). $v_{1-2}$ represents that the flow of information about a merger plan by the drivers of the merger is a factor; in this case, BG was reportedly aggressive throughout the contest. $\sigma_{1-2}$ reflects whether or not a consensus exists between the acquirer and target (e.g., to illustrate, the failure of the Big 3 automakers in the U.S. to build cars to suit its customers tastes rather than its own production interests reflects a lack of mutual consensus; Wall Street Journal, 2/9/07, p. A1,8). Mutual consensus did exist for BG-EOP, but apparently not for VR-EOP. After its initial bid was rebuffed, by proposing to substantially increase indebtedness and refusing to match BG's offer to make EOP's $8.4 billion non-secured bondholders whole, VR met substantial resistance among EOP's bondholders; i.e., proposing to push them down the default ladder was "Not welcome news for bondholders" per R. Haines of Credit Sights Ltd (Wall Street Journal, 2/2/07, p. C9). Although EOP's Directors had concluded that VR's bid remained too risky (www.nytimes.com, 2/6/07), BG countered with the winning bid of $55.50/share, an underbid to VR's bid.

Illusions appear to have played a factor on many levels and in driving the outcome. Bringing opposing views together helps to dispel the illusions (e.g., Lawless & Whitton, 2007). The two competitors (BG and VR) for EOP had competing visions, VR with the reputation of a superb commercial real estate manager and BG with the reputation of a sophisticated deal-maker. Their different pitches were designed to win over primarily their immediate (EOP Directors and stockholders) but also to a lesser degree their wider (stock market) audiences. The competition, we suggest, entangled both audiences (previously considering EOP to be too large, the real estate world was "stunned" that EOP could be taken over; Wall Street Journal, 2/2/07, p. C3). The opposing views of the drivers, expert though they may have been, become unreliable justifiers of their separate actions and beliefs (Shafir & LeBoeuf, 2002), as do members of the lesser informed target audiences who became engaged because of self-interest and because the bidding war was entertaining.[6] In addition, if the outcome is ultimately a success story, the bids indicated that a major transformation, but possibly illusory, occurred in the valuation of commercial real estate. It assumed that rents were increasing faster than they are, and that vacancy rates will continue to fall, a perception called "troubling" by M. Kirby of Green Street Advisors (www.nyt.com 2/3/07). It underscored the frenzy occurring that was driving the flow of capital into commercial rents.

There are other, more basic attempts that are relied upon to dispel illusion, such as business valuations. But Return On Investment (ROI) calculations often lead to business decisions to make no investment in R&D, leading Rouse and Boff (2003, p. 639) to conclude that decisions to invest in innovation are not based on traditional valuation techniques. However, as we suggested earlier, by using managed conflict such as internal debate, illusions may be dispelled by defining and justifying ROI in an organizational context. The use of expected utility by an organization produces internal debates and eventually justifications for an action that can make defensible business decisions (Bankes, 2002; Gelman et al., 2003; Rouse & Boff, 2003, p. 644). But ultimately, whether an illusion exists or not can only be decided by a test in the market (Kock, 2005). In addition, the risks associated with potential illusions can be mitigated with action; e.g., Blackstone Group reduced its potential risks by rapidly selling-off parts of its EOP acquisition after its deal was completed (www.nytimes.com, 2/6/07).

---

[6]Game theory is largely ineffectual here because its parochial view of two competitors does not easily address social welfare from the viewpoint of Nash equilibria (Lawless & Grayson, 2004); nor do the nominal values assigned to "cooperation" and "competition" in game theory work well: "the value of a strategy … only reflects … the expected value … [which may] introduce a higher risk" (Kock, 2005, p. 637) which can only be decided for strategies tested in the marketplace (p. 655); further, utility theory often does not reflect the human decision making that governs most business decisions (Rouse & Boff, 2003, p. 635).

**Future Research. Overview:**
We are following a dual computational track of developing quantum agents and constructing an organization composed of these bistable agents to form a social circuit and then to combine them into a computational representation of an organization to model and simulate (and eventually control with metrics) what we have and are finding in the field. In addition to computational approaches, we are working with two organizations to design and test metrics (i.e., MDRC and CBU). The fifth and final track is to continue with the development of theory.

**Future Research. Quantum agents**:
Our present research is directed at designing an organization composed of quantum (bistable) agents. These agents should be able to reside in at least one of two states: in a baseline or excited state; in an action or observational state; and under the influence of incommensurable belief illusion A or B. With roles as the anchors that build social structures like organizations, bistable agents under social psychological influences create a tension field between local and more global beliefs that is observable in democracies (information) but not in command decision-making or consensus-seeking systems, making democracies more successful than the latter in providing for the social welfare of their citizens but also less rational (Lawless et al., 2000).

Some of our previous research concerned the domain of modelling and implementing agent-based social simulations. Our goal has been to provide the best and most well-known existing tools with the ability to simulate complex societies through concurrent models. Recent tools, such as SWARM, REPAST, and MASON, allow only the execution of simulated concurrency, through a centralized scheduler. Recently, some members of our team have started a research project aiming to model and implement realistic concurrent simulations, using Petri Net formalisms and multi-threaded programming technologies (DIVERSITY, 2007). First results from this research, namely the extension of well-known agent-based simulation libraries, already allow the programming of truly concurrent agents (Louçã and Meneses, 2007). The ability to model and implement concurrent multi-agent simulations will be essential in the study of organizations composed of quantum (bistable) agents, were agents should be able to reside in at least one of two states in a concurrent manner. In this context, an interesting perspective in what concerns modelling quantum agents is using the language Z (Spivey, 2006) to formally describe the intricacy of interactions between complex bistable agents.

**Future Research. Conflict and Galois Lattices:**
We have been building agent-based, systems and Galois lattice (GL) models of organizations of our field and laboratory findings. A GL model may provide a logic structure to capture uncertainty in a social circuit. With humans, conflict and competition generate information and uncertainty and hold the attention of neutral observers who serve as judges. But with logic, building differential operators in symbolic models requires negation or ortho-complements that are difficult to locate in non-modular lattices. Indeed, Chaudron and his colleagues (2003) have proved that conjunctions of first order logic literals define a non-modular lattice (the cube model). The idea is to go back to elementary properties provided by negations, considered as a GL. We intend to upgrade such structures to enrich their capabilities to capture predicate logic properties including a conflict-adapted negation operator. If problem solving is cooperative, negation locates uncertainty at the point of least cooperation between opposing agents.

**Future Research. Field research (Medical Department Research Center--MDRC):**
Our evaluation of field data shows that the standing of MDRC within the Army research community could be improved by increasing its research productivity impact index by the:
1. Accurate capture of all scholarly products being produced by MDRC,
2. Encouragement to increase the number and quality of research protocols and those scholarly

    products produced by each protocol, and,
3. Continued application for external funding.

A system that effectively captures all aspects of the research process, from protocol submission and processing to publication of scholarly products or novel therapeutics will generate the highest quality data for productivity analysis and metric development. Based on field research, we believe this can best be achieved by developing an electronic protocol submission and management system with the capacity to generate real time metrics of productivity and quality (Lawless et al., 2006c). To achieve this end we intend to submit to the Army Medical leadership a business case analysis to fund this endeavor.

There are a number of commercial products available to meet some of these needs that address protocol submission and management. However, these products require modest customized re-engineering to permit metric tracking. On the other hand, a system could be developed that would process the necessary research documents and track productivity as well as provide a metric to assess the quality of research performed and publications from that research.

We have begun a process within MDRC that more accurately captures the scholarly products generated, which includes a publication clearance policy and internal education of investigators on the process. Currently, this is a paper-based process without the ability to track metrics. However, utilization of the system described above has the potential to facilitate this process immediately.

This system has the potential to be developed into DoD-wide electronic research data system with embedded metric tracking tools to accurately access organizational productivity and quality. Once these data are captured by such a system, research centers, including MDRC, could apply business tools in accord with our metrics, such as Lean Six Sigma, to identify problem areas, enable corrective measures, and initiate actions that would deliver the highest quality and beneficial research product for the taxpayer.

**Future Research. Metrics for a Central Business University's Graduate School (CBU):**

We have just begun to work with CBU's faculty administration to collect data to construct a metric for its faculty information reporting system (FIS). We will survey and then interview its faculty individually and in focus groups about the present but unfinished test version of its FIS reporting system. After the first stage of data collection is collected and analyzed, we will work to convert the initial test version of FIS into a release version. In the process, we will build, field test, and collect data on our online and real-time metrics to measure the performance of CBU against its regional peers.

**Future Research. The development of new theory:**

Dynamics introduced into our static model with coupled differential or discrete equations where managers of commercial businesses seek the center of gravity in a market (e.g., General Electric; www.nytimes.com, 1/20/07), or seek new markets as old ones mature or dry up (e.g., the ethanol revitalization of farm markets); or medical research managers seek more extramural funds or qualitatively better publications (e.g., MDRC medical research); (e.g., EU-MIS); or faculty leaders who seek to restructure their faculty with metrics to improve their performance versus their regional peers (e.g., CBU).

In our endeavors with the development of organizational theory, we will borrow from network theory to convert complex networks into series (AND) parallel (OR) circuits that can be solved, where we first expect to replicate that parallel networks reflect a tradeoff between complexity and productivity (Yu & Efstathiou, 2002). We also expect to replicate that increasing resources leads to less computational time within an organization, and that increasing the resources required ($\Delta R_r$) for a problem solution increases solution time (Reyck & Herroelen, 1999; also, we have found both of these effects in the field; Lawless & Whitton, 2007). After these replications, our most difficult challenge will be to construct and apply complex functions to the network.